\documentclass[preprint,preprintnumbers,amsmath,amssymb]{revtex4}
\usepackage{color}
\usepackage{graphicx}
\begin{document}
%\begin{CJK}[HL]{KS}{}
%=======================================================================================================
\title[Short Title]{Magnon-bandgap controllable artificial domain wall waveguide}
\author{\bf Hai Yu$^{1}$, Xiao-Ping Ma$^{2}$\footnote{Corresponding author E-mail: xpma1222@ybu.edu.cn}, Huanhuan Zhang$^{1}$, Xue-Feng Zhang$^{1}$, Zhaochu Luo$^{3}$, and Hong-Guang Piao$^{1,2}$\footnote{Corresponding author E-mail: hgpiao@ybu.edu.cn}, \\}
\affiliation{
$^1$ Hubei Engineering Research Center of Weak Magnetic-field Detection, China Three Gorges University, Yichang 443002,
P. R. China\\
$^2$ Department of Physics, Yanbian University,Yanji 133002, P. R. China\\
$^3$ School of Physics, Peiking University, Beijing 100871, P. R. China \\
}

%=======================================================================================================

\begin{abstract}
In this paper, a magnon-bandgap controllable artificial domain wall waveguide is proposed by means of micromagnetic simulation. By the investigation of the propagation behavior and dispersion relationship of spin waves in artificial domain wall waveguides, it is found that the nonreciprocal propagation of spin waves in the artificial domain walls are mainly affected by the local effective exchange field, and the magnon bandgap can be controlled by changing the maximum value of the effective exchange field.
In addition, it is observed that the artificial domain wall waveguides are structurally more stable than the natural domain wall waveguides under the same spin wave injection conditions, and the magnon bandgap of the artificial domain wall waveguides can be adjusted by its width and magnetic anisotropy parameters.
The bandgap controllable artificial domain wall scheme is beneficial to the miniaturization and integration of magnon devices and can be applied to future magnonic technology as a novel frequency filter.\par

%{\bf PACS number:} \ 75.30.Ds , \ 75.40.Gb, \ 75.78.Cd, \ 85.70.Kh \par
{\bf Key words:}\ spin waves, magnonics, magnonic waveguides, magnon filters
\end{abstract}\maketitle
%=======================================================================================================

As a collective precession behavior of spins in magnetic system\cite{magnon_Spintronics, Magnonic_filp}, the propagation of spin wave (SW) does not require the transfer of actual carrier, thus effectively avoiding thermal effects caused by carrier scattering (such as the electric current)\cite{Joule1, Joule2, Joule3}.
Moreover, the frequency of SW can reach THz level and its wavelength can also reach nanometer level\cite{ultrafast_stt, Long_propagation}, which is conducive to the miniaturization and integration of future spintronic devices. \par

The magnonics based on SW behaviors in nanometer scale has developed rapidly\cite{Source, review1, review2} in recent years, and various magnonic devices\cite{multiplexer, spinwave_analyzer, phase_sifter, Magnon_transistor, directional_coupler, majority_gate} have emerged as the times demand.
A key point for miniaturizing and integrating magnonic devices is to firstly find a magnonic waveguide that can stably propagate SWs and can easily control the direction of SW propagation.
For this purpose, K. Vogt, \textit{et al}. successfully modulated the direction of SW propagation in a curved Py microwire waveguide by using the Oersted field caused by pulsed electric current\cite{curved_wire}, which is an important experimental progress for the exploration of magnonic waveguides.
However, the fast attenuation of SW in the curved magnonic waveguide leads to short effective propagation path, which is not conducive to signal transmission.
Meanwhile, the modulation of Oersted field is not conducive to the miniaturization and integration of magnonic devices. \par

J. M. Winter predicted the possibility of SW propagating along a Bloch wall by theoretical calculation in 1961 \cite{theory_bloch_wall}, thus opening the new way to investigate whether the magnetic domain walls (DWs) can be used as magnonic waveguides.
Recently, the propagation behavior of SWs confined in the DWs has been demonstrated by micromagnetic simulation\cite{localized_sw_dw}, and it is believed that the DWs will have a good application prospect as magnonic waveguides.
It is interesting to note that when the Dzyaloshinskii-Moriya interaction (DMI)\cite{D_of_DMI, M_of_DMI} is introduced into the ferromagnetic thin film, the SW propagation behavior confined in the natural DWs exhibits nonreciprocity\cite{DMI_on_SW, theory_DMI}.
The simulation results show that when there is an interfacial DMI, the non-reciprocally propagated SWs in the N\'{e}el-type DW wall can achieve a higher group velocity at the same frequency, which provides a new method for DMI measurement\cite{sw_DW} .
Subsequently, K. Wagner, \textit{et al.}\cite{sw_dw_exm} experimentally observed the propagation of SWs confined in the natural DWs, further confirming the feasibility of magnonic waveguides based on the DWs.
Moreover, it has been found that the DW-based magnonic waveguide can be controlled by adjusting the natural DW structure, and an engineering reconfigurable magnonic waveguide scheme has been proposed \cite{reconfigurable_dw}.
Although the scheme has some potential application value\cite{dw_divce}, it is inevitably susceptible to receive external interference\cite{dw_move}, which will certainly have an adverse effect on the stability of magnonic devices.
Therefore, in order to realize the miniaturized low-power magnonic devices, it is necessary to develop a magnonic waveguide that can both stably propagate the SW and can modulate its propagation direction. \par

Thanks to the continuous development of the nanofabrication technology, various artificial spin textures such as artificial DWs\cite{Lou2,Lou4}, artificial Skyrmions\cite{A_skyrmion1, A_skyrmion2}, and so on, can be constructed at present.
Among them, the artificial DW has attracted much attention because it can be easily constructed in ferromagnetic thin films with perpendicular magnetic anisotropy (PMA) by simply changing the local magnetic magnetocrystalline anisotropy\cite{Lou4}.
Intuitively, SWs should also propagate in arti?cial DWs since it can propagate in natural DWs.
In this work, a novel magnonic waveguide scheme based on artificial DWs is proposed, and the feasibility of the scheme is demonstrated by micromagnetic simulation.
The results show that the artificial DW waveguide, like the natural DW waveguide, can also overcome the propagation limitation of the magnetic thin film for low-frequency SW modes while maintaining the nonreciprocal characteristics of SW propagation.
It is worth noting that the magnon bandgap of the artificial DW waveguides can be adjusted by the width and magnetocrystalline anisotropy of the artificial DW, which is very beneficial for the wide application of magnonic devices. \par

In order to investigate the SW propagation behavior in artificial DW waveguides, the micromagnetic simulation based on MuMax$^3$ software\cite{MuMax3} was performed in this work.
In all the simulations, the material parameters\cite{damping-PMA,DMI_of_CFB,DMI_of_CFB2} of CoFeB film with PMA are considered, the saturation magnetization $M_{s} = 1 \times 10^{6}$ A/m, the exchange stiffness coefficient $A = 15 \times 10^{-12}$ J/m, the PMA constant $K_{u} = 0 \sim 1 \times 10^{6}$ J/m$^{3}$, the interfacial DMI constant $D = 1.5$ mJ/m$^{2}$ and the Gilbert damping constant $\alpha$ is varied from 0.001 to 0.01.
The width of the CoFeB thin film is varied here from 205 nm to 860 nm while the length and thickness are fixed to be 2048 nm and 1 nm, respectively. The unit cell size is $1 \times 1 \times 1$ nm$^{3}$.
The artificial DWs are achieved initially by embedding local channels with in-plane magnetic anisotropy (IMA) in a PMA CoFeB film\cite{A_DW}. Due to the chiral coupling provided by DMI, a stable N\'{e}el-type artificial DW can be formed in the uniaxial IMA channel.
The SW is excited by a local sinusoidal magnetic field of $B = \textit{B$_0$}\sin(2\pi\textit{f} t)$ along the $x$ axis, where the field amplitude \textit{B$_0$} is set to be 100 mT, and the frequency \textit{f} varies from 0 GHz to 50 GHz. In addition, the width of artificial DW is varied in the range of $5 \sim 25$ \ nm to examine the channel width effect on SW propagation behavior. \par

In order to verify the feasibility and stability of magnonic waveguide scheme the structural stability of natural and artificial curved DWs was compared when a SW of \textit{f} = 20 GHz and B$_0$ = 100 mT was injected. The width of the artificial DW was set to be the same as that of the natural DW (5 nm), and a $90^{\circ}$ curved structure of both DW with 100-nm curvature radius was initially designed, as outlined by the dashed yellow lines in Fig. 1(a) and 1(b).
After magnetization relaxation, it is found that the curvature radius of the natural DW increases significantly whereas the artificial DW retained its original shape.
Moreover, after the 1-ns SW injection, the shape of the natural DW became distorted (see the solid and dashed orange lines in Fig. 1(c)), while the artificial DW retained it original shape during the SW propagation process, as shown in Fig. 1(c) and (d).
To further confirm the stability of the artificial DW, the difference of the magnetic domain area ($\Delta S$) with opposite magnetization orientation on both sides of the two kinds of DWs and the absolute value of normalized $z$-component magnetizations ($m_z$) over time were compared, as shown in Fig. 1(e).
It is found that the $\Delta S$ and $m_z$ of the natural DW are obviously changed during the SW injection but not for the artificial DW case (the artificial DW is very stable) implying that the magnonic waveguide scheme based on artificial DWs is feasible.\par

To systematically explore the magnonic waveguide properties of the N\'{e}el-type artificial DW, the propagation characteristics of SWs in an artificial DW with $d=20$ nm and $K_u=0$ J/m$^3$ were investigated by changing the excitation frequency.
As shown in Fig. 2(a), the SW cannot form and propagate in the artificial DW (see the inset of Fig. 2(a)) under 4 GHz low-frequency excitation. However, when the frequency is increased to 14 GHz, a nonreciprocal propagating SW behavior can be clearly observed in the artificial DW, and there is no excitation of SWs in the domains on either side of the artificial DW, as shown in Fig. 2(b).
It is very similar to the propagation behavior of low-frequency SWs in natural DWs, which indicates that artificial DWs have magnonic waveguide properties similar to natural DWs.
Moreover, when the frequency is further increased to 30 GHz, SWs can be seen in both the magnetic domain and the artificial DW regions, as shown in Fig. 2(c).
This result implies that the artificial DW has a bandgap for SW modes that differs significantly from the gapless natural DW\cite{sw_DW}, but the bandgap width is obviously smaller than that of the magnetic domain.
In order to further confirm that the artificial DW has a smaller bandgap, the dispersion relationships of the SW propagating in the magnetic domain (purple squares), artificial DW (orange circles) and natural DW (green triangles) are compared, as shown in Fig. 2(d).
It is worth noting that the magnon bandgap of artificial DW (\textit{f$_0$}) lies between that of the domain and natural DW, and the artificial DW waveguide has asymmetric dispersion (nonreciprocity) like the natural DW waveguide.
The SW dispersion relationship in artificial DW is the same as that of natural DW and follows the exchange-dominated SW mode \cite{sw_DW},
\begin{equation}
f=\frac{\gamma}{2M_{s}\pi}\sqrt{2Ak_{x}^{2}(2Ak_{x}^{2}-\frac{2\mu_{0}M_{s}^{2}t_{h}}{t_{h}+d\pi}+\frac{D\pi}{2d})}+\frac{D\gamma}{4M_{s}}k_{x}+f_0.
\end{equation}
Where, $f$ is the SW frequency propagating along the artificial DW, $\gamma$ is the gyromagnetic ratio, $M_{S}$ is the saturation magnetization, $A$ is the exchange stiffness, $k_x$ is the wave vector of SW propagating along the artificial DW, $\mu_{0}$ is the vacuum permeability, $t_{h}$ is the film thickness, $d$ is the width of artificial DW, $D$ is the interface DMI intensity, and $f_{0}$ represents the SW frequency for $k=0$ in artificial DW.
This is an interesting discovery that reminds us of semiconductors, which have been widely used in this past, and have tunable energy band between conductors and insulators.\par

In order to find the direct cause of the magnon band shift in the artificial DWs, the distributions of the effective anisotropy field ($B_{anis}$), demagnetization field ($B_{demag}$) and exchange field ($B_{ex}$) along the $y$-axis transversals of the artificial DW are analyzed, as shown in Fig. 3(a-c).
It is generally believed that the gapless of SW modes in the natural DWs is due to the $B_{anis}$ associated with the PMA is canceled out in the DW center\cite{sw_DW}.
From the Fig. 3(a), it is found that the $B_{anis}$ distributions of artificial DW and natural DW are quite different, the effective field in both walls is close to 0, which obviously does not explain the $f_0$ shift.
Similarly, the effective $B_{demag}$ field inside both DWs is also close to 0 (see Fig.3 (b)), which again does not explain the $f_0$ shift.
However, it is interesting to find that there are significant differences in the distribution of the effective $B_{ex}$ fields inside the two DWs, as shown in the Fig. 3(c).
It is clearly observed that there are significant differences in the peaks of the $B_{ex}$ fields at the two edges of both DWs, and it can be preliminarily concluded that $B_{ex}$ is the real cause of the $f_0$ shift in the artificial DWs.
It is well known that the formation and propagation of SWs in magnetic systems originate from the exchange interactions between their internal spins.
Therefore, we firmly believe that the maximum value of the $B_{ex}$ inside the DWs is the key factor that controls the magnon bandgap.\par

To further verify the above conclusion, the $f_0$ shift is investigated by changing the DW width $d$, internal IMA parameter $K_u$ and damping constant $\alpha$ in the artificial DW region.
As shown in Fig. 3(d), it can be clearly observed that the variation trend of the maximum value of $B_{ex}$  ($B_{ex}^{max}$, green solid circles) with the $d$ is basically consistent with the trend of $f_0$ shift (hollow symbols), while the trend of the minimum values ($B_{demag}^{min}$ and $B_{anis}^{min}$) of $B_{demag}$ (solid squares) and $B_{anis}$ (solid triangles) is quite different.
Moreover, as shown in Fig. 3(e), with the change of $K_u$, it can be clearly observed that the change trend of the $B_{ex}^{max}$ (green solid circles) is linearly correlated with $K_u$, just like the trend of $f_0$ shift (hollow symbols), which further implies that the $B_{ex}$ dominates the SW propagation behavior in artificial DW.
It is worth mentioning that the variation trend of $B_{demag}^{min}$, $B_{anis}^{min}$, and $B_{ex}^{min}$ values with $K_u$ and $d$ is inconsistent with the $f_0$ shift trend (not shown in here).
In addition, it can be seen from the Fig. 3(d) and 3(e) that the change of $\alpha$ hardly affects the variation trend of $f_0$ shift with $d$ and $K_u$.
Moreover, it can be seen from Fig. 3(d) that the effect of the $d$ on the $f_0$ shift is limited, and its effective range is less than $\sim 20$ nm.
Since the exchange length $l_{ex}=\sqrt{2 A_{ex} / (\mu_0 M_S^2)}$ in the artificial DW equals $\sim 4.9 $ nm, which means that the $B_{ex}^{max}$ in the artificial DW reaches the maximum limit value ($\sim 0.58$ T, see the dotted green line in Fig. 3(c)) when $d \geq 4 l_{ex} \approx 19.6$ nm.
According to the Bloch wall width definition, the limit width of a single artificial DW will be equal to $4 l_{ex}$, which means that when $d$ exceeds $4 l_{ex}$, two separated natural DWs might be formed within the artificial DW depending on the exchange energy barrier $E_{ex}= -\frac{1}{2} \vec{M} \cdot \vec{B}_{ex} $ at the two edges of the artificial DW, as  shown in Fig. 3(c).

As the dotted blue line shown in Fig. 3(c), two completely separated natural DWs formed at the two edges of the 40-nm-wide artificial DW can be clearly observed.
However, due to the dynamic dipolar coupling\cite{dw_divce} between adjacent quasi-natural DWs in the artificial DW, the effective width of the artificial DW as a magnon waveguide can be further widened, which will certainly benefit its wide application.
According to the above discussion, it is certain that in the artificial DW with discontinuous material parameters, the magnon bandgap is mainly determined by the effective $B_{ex}^{max}$ in the artificial DW, which provides controllability for its magnonic waveguide applications.

To demonstrate the application prospect of artificial DW in future magnonic devices, a nonreciprocal magnonic waveguide with controllable magnon bandgap is designed, as shown in Fig. 4.
Firstly, a frequency filtering function of magnon waveguide whose band gap can be controlled by artificial DW width is demonstrated, as shown in Fig. 4 (a).
For simplicity, the anisotropic parameter $K_u$ is set to zero here.
It is found that the SW with a frequency of 11 GHz can only forms and nonreciprocally propagates in the $d=5$-nm artificial DW, but not in the $d=10$-nm artificial DW, which is attributed to the bandgap difference between different width artificial DWs.
In the case of $d = 10$-nm artificial DW, the $f_0 = 11.3$ GHz, which is grater than 11 GHz, while for $d = 5$-nm case, the $f_0 = 9.5$ GHz, which is less than 11 GHz, as shown in Fig. 3(e).
Although the frequency filtering function of magnon waveguide is successfully realized based on the artificial DW width, its modulated frequency band range for SW is narrow, and its application range is bound to be limited.
Therefore, based on the linear relationship between $f_0$ and $K_u$ (see Fig. 3(f)), a magnon-bandgap controllable waveguide scheme is proposed based on anisotropic parameter of artificial DW.\par

In order to realize the frequency filtering function of the magnon-bandgap controllable waveguide in a wide frequency range, $d=20$ nm artificial DWs with three different $K_u$ are designed side by side on a PMA film at 200 nm spacing, as shown in the Fig. 4(b)-(d).
Here, the IMA parameters in three artificial DWs are $K_{u0} = 0$ J/m$^{3}$, $K_{u1} = 1 \times 10^{5}$ J/m$^{3}$, and $K_{u2} = 2 \times 10^{5}$ J/m$^{3}$, respectively.
It can be clearly observed from Fig. 4(b) that when the source excitation frequency is 12.5 GHz, a SW only forms in the artificial DW of $K_{u0}$ and propagates nonreciprocally along the domain walls, while it do not form in the other two DWs.
As expected the source excitation frequency is changed to 16 GHz, the SWs can form and nonreciprocally propagates in the artificial DWs of $K_{u0}$ and $K_{u1}$, but not in the DW of $K_{u2}$.
When the source excitation frequency is increased to 19.5 GHz, SWs are formed in all of the three artificial DWs and corresponding nonreciprocal propagation behavior appears as well.
As demonstrated above, a magnon-bandgap controllable waveguide can be achieved using artificial DWs, which is very useful for the development of stable magnonic devices.

\par

%=======================================================================================================
In conclusion, a magnon-bandgap controllable artificial DW waveguide is proposed in this work, whose structural properties are more stable than those of natural DW waveguides, allowing for the miniaturization and integration of future magnonic devices.
Moreover, due to the controllability of the magnon bandgap of the artificial DW waveguides, it is expected to be used as a frequency filter of SWs in the development of magnon-related technologies.
\par

This work was supported by the National Natural Science Foundation of China (Nos. 52271160), the Yichang Natural Science Research Project (No. A22-3-010) and the Yanbian University Research Project (Grant No. 482022104).
This work was also supported by the Yichang Key Laboratory of Magnetic Functional Materials.
Thanks to Dr. Jiangwei Liu from CTGU for polishing this article in English. \par

\textbf{DATA AVAILABILITY}\\
The data that support the findings of this study are available from the corresponding author upon reasonable request.

\newpage

%=================================================================================================

\newpage

%======================================================================================================
{\centering FIGURES\\[1cm]}

FIG.1. The structural stability of curved waveguides for natural DW and artificial DW is compared under the same SW injection.
(a) and (b) show the initial profiles of the curved natural and artificial DWs, respectively. (c) and (d) show the profiles of the curved natural and artificial DWs after the 1-ns SW injection, respectively.
The red line represents the SW source position, and the $\Delta S$ represents the structural deformation after the 1-ns SW injection. The gradient black bar and color bar indicate the $m_z$ and $\Delta m_z$ distributions in the CoFeB films, respectively.
(e) shows the variations of $m_z$ (dotted lines) and the difference of the $m_z \geq 0.9$ and $m_z \leq -0.9$ areas $\Delta S$ (solid lines) over time in two films containing natural and artificial DWs.
\\[1cm]

%======================================================================================================
FIG.2. The propagation of the (a) 4-GHz, (b) 14-GHz, and (c) 30-GHz SWs in the CoFeB film containing artificial DW $d = 20$ nm (the area between two adjacent dotted red lines), and (d) the dispersion relations of SW propagating in uniform magnetic domain, natural DW, and artificial DW with $d = 20$ nm.
The inset of (a) schematically shows a side view of the magnetization distribution in the CoFeB film.
The black lines in (a)-(c) represent the SW source positions, and the color bar indicate the $\Delta m_z$ distributions in the CoFeB films.
\\[1cm]

%======================================================================================================
FIG.3. The side views of the (a) anisotropy field ($B_{anis}$), (b) demagnetization field ($B_{demag}$), and (c) exchange field ($B_{ex}$) distributions near the natural and the artificial DWs with $d = 20$ nm along the $y$-axis. The light-gray and blue areas indicate the width of the natural and artificial DWs.
The dotted blue line and yellow area graph in the (c) indicate the $B_{ex}$ and the $E_{ex}$ (exchange energy) distributions for the artificial DW with $d = 40$ nm, respectively.
(d) shows the effect of $d$ and $\alpha$ on the $f_0$, and the change of the $B_{ex}^{max}$, $B_{demag}^{min}$, and $B_{anis}^{min}$ with $d$.
(e) shows the effect of $K_u$ and $\alpha$ on the $f_0$, and the change of the $B_{ex}^{max}$ in the $d = 20$ nm artificial DW with $K_u$.
\\[1cm]

%=====================================================================================================
FIG.4. (a) shows the SW formations in artificial DW of $d = 5$ nm and $d = 10$ nm under 11-GHz excitation of the SW source. The inset schematically shows a side view of magnetization distribution in the CoFeB film containing artificial DWs of different widths.
In the CoFeB film contains three $d=20$-nm artificial DWs with anisotropy constants $K_{u0} = 0$ J/m$^{3}$, $K_{u1} = 1 \times 10^{5}$ J/m$^{3}$, and $K_{u2} = 2 \times 10^{5}$ J/m$^3$, the formation of SWs under (b) 12.5 GHz, (c) 16 GHz excitations, and (d) 19.5 GHz is shown, respectively.
The inset schematically shows a side view of magnetization distribution in the CoFeB film containing artificial DWs of different $K_{u}$. The black lines represent the SW source positions, and the color bar indicates the $\Delta m_z$ distributions in the CoFeB films.
\\[1cm]

%======================================================================================================
\newpage
\begin{figure}[!h]\label{f001}
\setlength{\abovecaptionskip}{10pt}\centering
\includegraphics[width=1\textwidth]{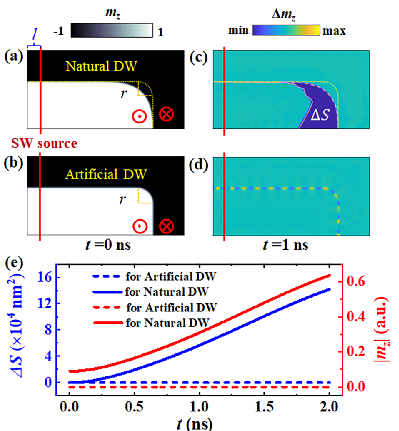}\\FIG. 1
%\caption{}
\end{figure}\newpage
%======================================================================================================
\newpage
\begin{figure}[!h]\label{f001}
\setlength{\abovecaptionskip}{10pt}\centering
\includegraphics[width=1\textwidth]{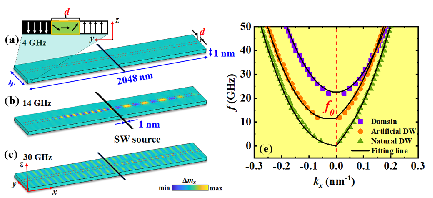}\\FIG. 2
%\caption{}
\end{figure}\newpage
%======================================================================================================
\newpage
\begin{figure}[!h]\label{f001}
\setlength{\abovecaptionskip}{10pt}\centering
\includegraphics[width=1\textwidth]{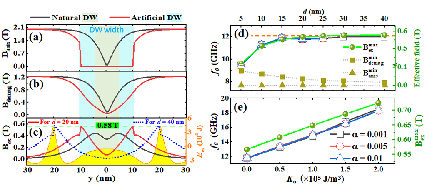}\\FIG. 3
%\caption{}
\end{figure}\newpage
%======================================================================================================
\newpage
\begin{figure}[!h]\label{f001}
\setlength{\abovecaptionskip}{10pt}\centering
\includegraphics[width=1\textwidth]{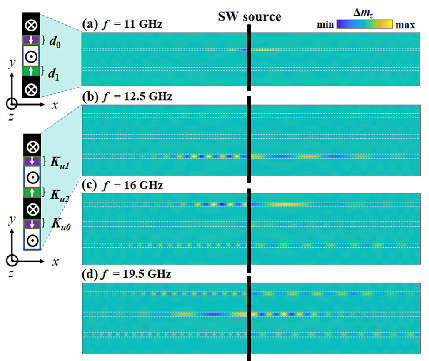}\\FIG. 4
%\caption{}
\end{figure}\newpage
%======================================================================================================

%\end{CJK}
\end{document}